\documentclass[aps,prl,reprint,showpacs,superscriptaddress,longbibliography]{revtex4-1}
\usepackage[english]{babel}
\usepackage{amsmath,amssymb,bbm,graphicx,color,comment,txfonts}
\usepackage[bookmarks=true,colorlinks,citecolor=blue,urlcolor=blue]{hyperref}
\usepackage{dsfont}
\usepackage{soul}

\usepackage{bbm}
\usepackage{float}

\usepackage{dcolumn}   
\usepackage{bm}        
\usepackage{filecontents}
\usepackage{lineno}

\usepackage{mathtools}
\usepackage[usenames,dvipsnames]{xcolor} 

\usepackage[export]{adjustbox}
\usepackage{adjustbox}

\usepackage{braket}
\usepackage{tikz}
\usepackage{bbold}

\newcommand{\bs}[1]{\boldsymbol{#1}}

\let\baraccent=\= \renewcommand{\=}[1]{\stackrel{#1}{=}}
\newcommand{\eq}[1]{Eq.\thinspace(\ref{#1})}
\newcommand{\fig}[1]{Fig.\thinspace{}\ref{#1}}
\newcommand{\fc}[1]{({#1})}
\newcommand{\figc}[2]{Fig.\thinspace{}\ref{#1}\thinspace{}\fc{#2}}

\usepackage[normalem]{ulem}

\begin{document}
\title{Anomalous Diffusion in Dipole- and Higher-Moment Conserving Systems}
\author{Johannes Feldmeier}
\affiliation{Department of Physics and Institute for Advanced Study, Technical University of Munich, 85748 Garching, Germany}
\affiliation{Munich Center for Quantum Science and Technology (MCQST), Schellingstr. 4, D-80799 M{\"u}nchen, Germany}

\author{Pablo Sala}
\affiliation{Department of Physics and Institute for Advanced Study, Technical University of Munich, 85748 Garching, Germany}
\affiliation{Munich Center for Quantum Science and Technology (MCQST), Schellingstr. 4, D-80799 M{\"u}nchen, Germany}

\author{Giuseppe De Tomasi}
\affiliation{T.C.M. Group, Cavendish Laboratory, JJ Thomson Avenue, Cambridge CB3 0HE, United Kingdom}

\author{Frank Pollmann}
\affiliation{Department of Physics and Institute for Advanced Study, Technical University of Munich, 85748 Garching, Germany}
\affiliation{Munich Center for Quantum Science and Technology (MCQST), Schellingstr. 4, D-80799 M{\"u}nchen, Germany}

\author{Michael Knap}
\affiliation{Department of Physics and Institute for Advanced Study, Technical University of Munich, 85748 Garching, Germany}
\affiliation{Munich Center for Quantum Science and Technology (MCQST), Schellingstr. 4, D-80799 M{\"u}nchen, Germany}
\date{\today}

\begin{abstract}
The presence of global conserved quantities in interacting systems generically leads to diffusive transport at late times. Here, we show that systems conserving the dipole moment of an associated global charge, or even higher moment generalizations thereof, escape this scenario, displaying subdiffusive decay instead. Modelling the time evolution as cellular automata for specific cases of dipole- and quadrupole-conservation, we numerically find distinct anomalous exponents of the late time relaxation. We explain these findings by analytically constructing a general hydrodynamic model that results in a series of exponents depending on the number of conserved moments, yielding an accurate description of the scaling form of charge correlation functions. We analyze the spatial profile of the correlations and discuss potential experimentally relevant signatures of higher moment conservation.
\end{abstract}

\maketitle

\textit{\textbf{Introduction.}}-- Thermal equilibrium of generic systems is characterized by a finite number of conserved quantities, such as energy, particle number or charge. A coarse grained nonequilibrium time evolution that irreversibly leads to such an equilibrium state is therefore expected to be dominated by transport of these quantities at late times, smoothing out inhomogeneities of the initial state~\cite{chaikin_lubensky_1995}.
This framework also extends to closed, interacting quantum systems, where the corresponding time scale of transport, separating `early' and `late' times, is usually marked by the onset of local thermalization. A phenomenological description of the ensuing dynamics can then be given in terms of a classical hydrodynamic description, with quantum properties merely entering the effective diffusion constant~\cite{Mukerjee06, Lux14, Bohrdt16,leviatan2017quantum, Rakovszky18,Khemani18Hyd, Parker19, Gopalakrishnan19, Schuckert20}.

A particularly intriguing set of conserved quantities arises in the presence of a global $U(1)$ charge together with the conservation of one or several of its higher moments, such as dipole- or quadrupole-moment. These \textit{higher-moment conserving models} have attracted much attention in the context of fractons~\cite{Chamon05,Vijay15,PretkoSub,williamson2019_fractonic,Bulmash18b,Gromov19Mul, you2019fractonic, pretko2020fracton}, and are realizable in synthetic quantum matter~\cite{Guardado20} or solid state systems~\cite{you2019_fracton,yan2020_pyrochlore}.
In particular, the intertwined relation between the internal charge conservation law and its dipole moment has recently been shown to have significant impact on nonequilibrium properties~\cite{Pai18, Sala19, Vedika19, Sanjay19,Taylor_19, khemani20192d,Sanjay19b, Rakovszky20}. In the most severe cases, for short-ranged interactions and low spin representations, the system fails to thermalize due to a strong fragmentation of the many-body Hilbert space into exponentially many disconnected sectors~\cite{Sala19, Vedika19}, giving rise to statistically localized integrals of motion~\cite{Rakovszky20}.

In this work, we study late time transport in ergodic models of dipole- and even higher moment conserving 1D systems. We avoid strong Hilbert space fragmentation by including longer range interactions and higher spin representations. Building on the intuition of an effective classical description in the regime of incoherent transport, we employ a cellular automaton approach to numerically study the long time dynamics (see e.g. \cite{Medenjak17, Gopalakrishnan_2018, Iaconis19} for related approaches). 
We find anomalously slow, subdiffusive transport of the underlying charges, described by a cascade of exponents depending on the highest conserved moment.
We further develop a general analytic hydrodynamic approach, valid for arbitrary  conserved multipole moments that is in full agreement with our numerical results. Moreover, we discuss experimental characteristics of higher-moment conservation and the consistency of our findings with quantum dynamics.\\

\begin{figure}[t]
\centering
\includegraphics[trim={0cm 0cm 0cm 0cm},clip,width=0.99\linewidth]{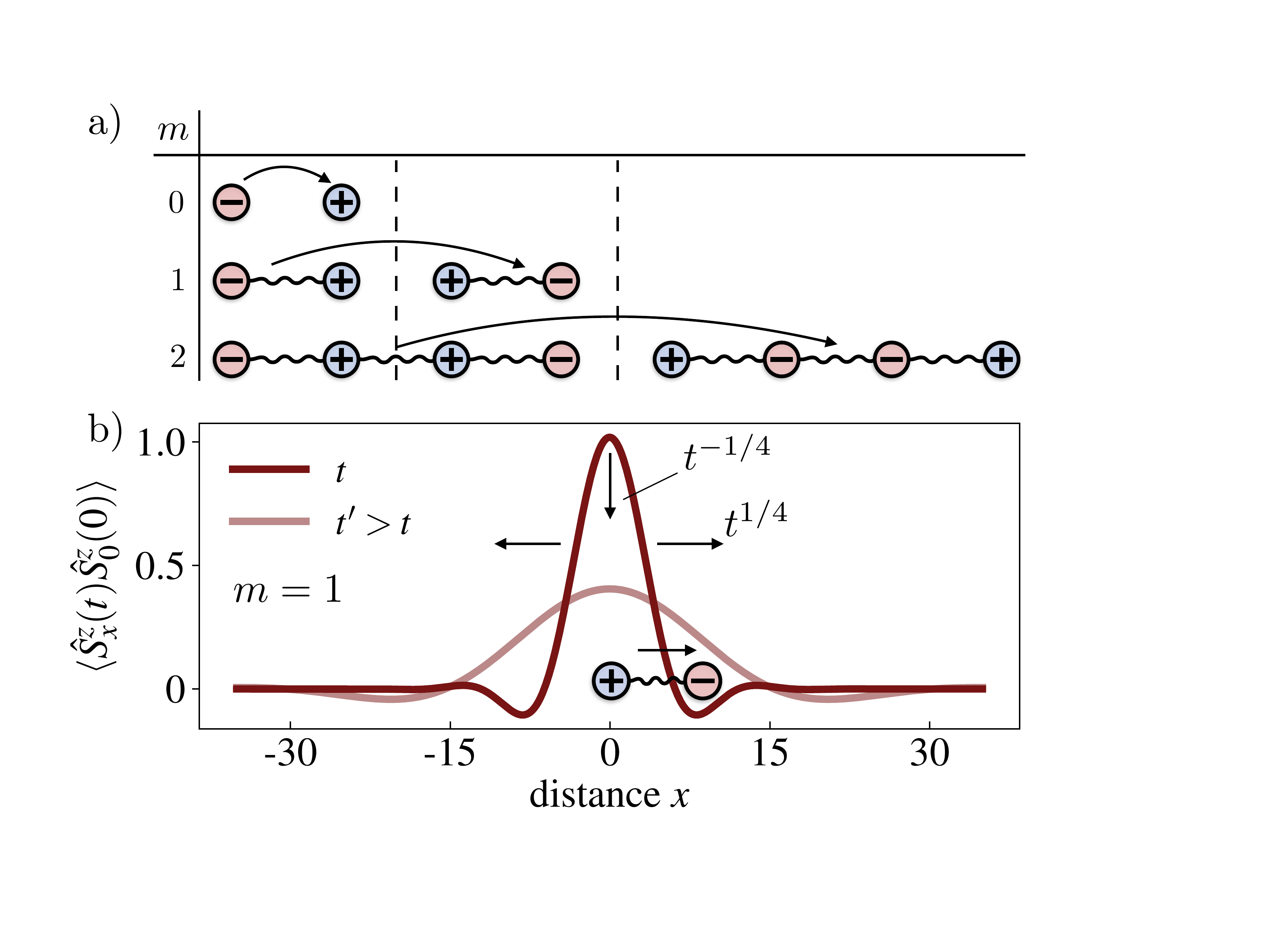}
\caption{\textbf{Higher Moment Conservation Laws}. \textbf{a)} Recursive construction: A charge-conserving move ($m=0$) creates a local dipole, which in turn is a charge-neutral dynamical object of a dipole-conserving model ($m=1$). This process is iterated to conserve higher moments. \textbf{b)} The late time dynamics of charges exhibits subdiffusive decay, with algebraic exponents depending on the highest conserved charge moment (here: $m=1$).}
\label{fig:1}
\end{figure}

\textit{\textbf{Higher-Moment Conserving Models.}}--
We start by constructing generic Hamiltonians conserving arbitrary moments of the charge. These will serve as input for the definition of suitable automata dynamics, as well as the derivation of an effective hydrodynamic description.
The construction of such models is best understood recursively, starting from a simple Hamiltonian of the form $\hat{H} = \hat{H}^{(0)}_{r=2} + \hat{H}_z$, with $\hat{H}^{(0)}_{r=2}=\sum_x(\hat{S}^+_x\hat{S}^-_{x+1}+h.c.)$ hosting local XY-type terms of range $r=2$ that conserve the total charge $Q^{(0)}=\sum_x\hat{S}^z_x$, and $\hat{H}_z$ containing arbitrary local terms diagonal in the $\hat{S}^z$-basis that render the model non-integrable. The $\hat{S}^\pm_x$, $\hat{S}^z_x$ are spin operators in a given representation $S$. Here, an elementary term $h^{(0)}_{2}(x)\equiv \hat{S}^+_x\hat{S}^-_{x+1}$ can be interpreted as the creation of a dipole against some background. A new term that additionally conserves the dipole moment $Q^{(1)}=\sum_x x\,\hat{S}^z_x$ can then be obtained by simply multiplying this operator with its hermitian conjugate at some shifted position, e.g. $h^{(1)}_3(x)=\left(h^{(0)}_{2}(x)\right)^\dagger h^{(0)}_{2}(x+1)$, yielding
$\hat{H}^{(1)}_3 = \sum_x \hat{S}^-_x\left(\hat{S}^+_{x+1}\right)^2 \hat{S}^-_{x+2}+h.c.$, a model that has been studied in the context of Hilbert space fragmentation~\cite{Sala19, Vedika19, Sanjay19,Taylor_19, khemani20192d,Sanjay19b, Rakovszky20}.

\begin{figure*}[t!]
\centering
\includegraphics[trim={0cm 0cm 0cm 0cm},clip,width=0.99\linewidth]{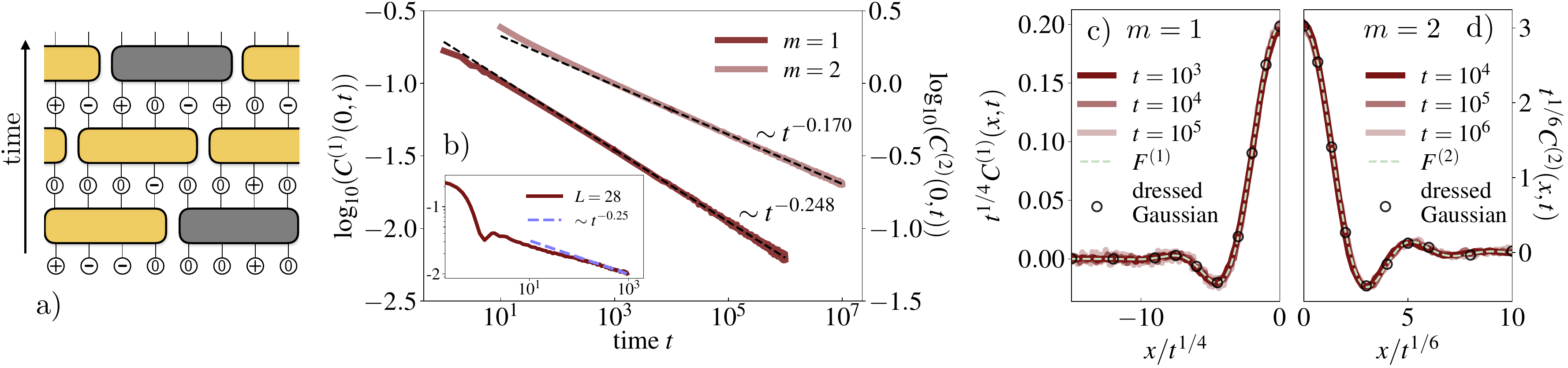}
\caption{\textbf{Hydrodynamics.} \textbf{a)} Illustration of an automaton circuit, for $m=1$ using dipole-conserving updates of range four between product states (spin representation $S=1$). With some finite probability, updates are either applied (yellow gates) not (grey gates), yielding effectively stochastic updates. \textbf{b)} Return probability $C^{(m)}(0,t)$ for dipole- and quadrupole conservation. The long time behavior approaches an $m$-dependent algebraic decay $\sim t^{-1/2(m+1)}$. The numerical values of the exponents where extracted from fits over the latter three time decades (dashed lines). Inset: Return probability of a small system size, dipole-conserving quantum model (spin $S=1/2$), consistent with subdiffusive decay. \textbf{c)+d)} Scaling collapse for $m=1$ and $m=2$ according to the long wave-length description \eq{eq:3.2}. In addition to the numerical data, the fundamental solution of \eq{eq:3.2} (dashed line) and the corresponding Gaussian expansion (see main text) up to order $n=4$ (circles) are shown. The system size is $L=10^4$ and correlations were averaged over at least $10^3$ random initial states in all panels.}
\label{fig:2}
\end{figure*}

The above recursion can be iterated to obtain models conserving arbitrary moments of the charge
\begin{equation} \label{eq:1}
Q^{(m)}=\sum_x x^m\,\hat{S}^z_x.
\end{equation}
Formally, we consider an $m^{th}$ moment conserving Hamiltonian of range $r$ in the form $\hat{H}^{(m)}_r=\sum_x h^{(m)}_r(x)+h.c.$, whose local terms can be expanded as
\begin{equation} \label{eq:1.1}
h^{(m)}_r(x) = \bigotimes_{i=0}^{r-1} \left(\hat{S}^{\mathrm{sgn}[\sigma_m(i)]}_{x+i}\right)^{|\sigma_m(i)|}, \mathrm{with}\;\,  \sigma_m(i) \in \mathbb{Z},
\end{equation}
where by definition $\sigma_m(0)\neq 0$, $\sigma_m(r-1)\neq 0$, and $\mathrm{sgn}[\cdot] \in \{+,-\}$ is the signum function. For the $XY$-terms, $\sigma_0(0)=-\sigma_0(1)=1$. Again, arbitrary terms diagonal in $\hat{S}^z$ could be added to \eq{eq:1.1} without affecting the conservation laws.
Analogous to the argument above, given $h^{(m-1)}_r(x)$, we can then construct a $(r+\ell)$-range term that additionally conserves the $m^{th}$ moment by imposing the recursive relation
\begin{equation} \label{eq:1.2}
\sigma_{m}(i) = -\sigma_{m-1}(i) + \sigma_{m-1}(i-\ell),
\end{equation}
on the exponents of the spin ladder operators. \eq{eq:1.2} reflects the construction of $\hat{H}^{(m)}_{r+\ell}$ via shifting an elementary \emph{$m$-pole} by $\ell$ sites. As illustrated in \figc{fig:1}{a}, the elementary \emph{$m$-pole} configurations have vanishing lower moments and a spatially independent $m^{th}$ moment, similar to usual charges. However, their number is not conserved.

We notice that \eq{eq:1.2} can be rephrased as a discrete lattice derivative of spacing $\ell$, $\sigma_{m}(i) = -\Delta_x \left[\sigma_{m-1}\right](i)$, which implies
$\sigma_{m}(i) =(-\Delta_x)^m[\sigma_0](i).$
If the elementary $XY$-terms are interpreted as a finite difference with spacing $\ell=1$, $(-\Delta_x)[f](0)=\sum_i \sigma_0(i)\, f(i)$ with some lattice function $f(i)$, the exponents $\sigma_m(i)$ effectively correspond to a lattice discretization of the $(m+1)^{st}$ derivative
\begin{equation} \label{eq:1.3}
(-\Delta_x)^{m+1}[f](0) = \sum_i \sigma_m(i)\, f(i).
\end{equation} 
Using the spin commutation relations and \eq{eq:1.3}, we see that $[Q^{(n)},h^{(m)}_{r}(x)] = \sum_i \sigma_m(i)\, (x+i)^n = (-\Delta_x)^{m+1}[x^n] = 0$ for $n\leq m$, i.e. all moments $Q^{(n\leq m)}$ are indeed conserved. The same holds for longer range Hamiltonians, using alternative discretization schemes of the involved derivatives. We note that this is a discretized version of the field theory construction in Ref.\cite{Gromov19Mul}.\\

\textit{\textbf{Cellular Automaton Approach.}}--
Studying the full quantum evolution of \eq{eq:1.1} is challenging. However, we can make progress by considering a classical automaton time evolution that respects the same conservation laws of \eq{eq:1}, which are the crucial properties concerning the late time dynamics. The discrete automaton evolution consists of a sequential application of local updates mapping $z$-basis product states onto $z$-basis product states. The updates are designed to mimick the action of the Hamiltonian by updating between local strings of spins $\bs{s}(x)=(s_x,...,s_{x+r-1})$, with $s_i\in\{-S,...,S\}$, that are connected by the local terms of \eq{eq:1.1} and thus feature the same conserved charge moments, similar to Refs.~\cite{Medenjak17, Iaconis19}. The time evolution can then be represented by a classically simulable circuit, see \figc{fig:2}{a}. Furthermore, by assigning a finite acceptance probability to each update, we simulate a \textit{stochastic} automaton~\cite{Ritort03}, enabling us to meaningfully study dynamics starting from fixed initial states. More details can be found in the supplemental material. We emphasize that the details of the implementation, including its stochastic nature, are \textit{not} essential to the hydrodynamics studied in the following.\\

\textit{\textbf{Hydrodynamics.}}--
To understand the dynamics of charges at late times, the main quantity studied within our numerical approach are the (infinite temperature) correlation functions 
\begin{equation} \label{eq:3.3}
C^{(m)}(x,t)=\braket{\hat{S}^z_x(t)\hat{S}^z_0(0)},
\end{equation}
and particularly the return probability $C^{(m)}(0,t)$, where $\braket{...}$ denotes an average of $\hat{S}^z_x(t)\hat{S}^z_0(0)$ over randomly chosen initial states of the automaton dynamics. For concreteness, we study a dipole conserving model with $S=1$ including interaction terms of \eq{eq:1.1} of ranges $r=3$ and $r=4$~\footnote{This model was shown to thermalize for typical initial states in Ref.~\cite{Sala19}.}, as well as a quadrupole conserving model with larger spin representation $S=4$ including ranges $r=4$ and $r=5$, both on system sizes up to $L=10^4$. While generally the hydrodynamic tails for given $m$ are expected to be universal for sufficiently ergodic systems, in numerical practice, larger $r$ and $S$ will provide faster convergence to this long time behavior. As the dynamics is expected to become slower upon increasing $m$, i.e. the number of constraints, we choose a larger spin representation $S=4$ in the quadrupole case to allow for a more accurate determination of algebraic exponents.

We show the results of the automaton evolution for the return probability $C^{(m)}(0,t)$ in \figc{fig:2}{b}. In contrast to ordinary diffusive systems which possess the scaling $C^{(0)}(0,t)\sim t^{-1/2}$~\cite{chaikin_lubensky_1995}, we numerically estimate the algebraic late time decay exponents to be $C^{(1)}(0,t)\sim t^{-0.248}\approx t^{-1/4}$ in the dipole-conserving model, and $C^{(2)}(0,t)\sim t^{-0.170}\approx t^{-1/6}$ for the conservation of quadrupole moment. Consistent with these results, we observe a subdiffusive decay of $C^{(1)}(0,t)$ for a dipole-conserving $S=1/2$ quantum spin chain as shown in the inset of \figc{fig:2}{b} (see the supplemental material for more details on the quantum case). This signals significant deviations from generic diffusion, induced by the higher-moment conservation laws \eq{eq:1}. \\

To understand how this slow anomalous diffusion can emerge from a classical hydrodynamic description of a quantum evolution, we consider the Heisenberg evolution equation of the charge density $\hat{S}^z_x$ for the previously introduced models. This yields $\frac{d}{dt}\hat{S}_x^z=\frac{i}{\hbar}[\hat{H}^{(m)}_r,\hat{S}_x^z]=(-\Delta_x)^{m+1}\Omega^{(m)}_x$ using \eq{eq:1.3}, with $\Omega^{(m)}_x=-\frac{i}{\hbar}\bigl(h^{(m)}_r(x)-h.c.\bigr)$, which takes the form of a generalized `multipole current' of $m$-poles (In fact, this is a one-dimensional version of generalized currents appearing in fractonic systems~\cite{yuan2020_superfluid}). This form of the time evolution applies to arbitrary Hamiltonians conserving the $m^{th}$ moment of the charge, with microscopic details only entering $\Omega^{(m)}_x$. Imposing a continuity equation for the charge density, $\frac{d}{dt}\hat{S}_x^z=(-\Delta_x)J^{(m)}_x,$
we obtain the form of the charge current $J^{(m)}_x=(-\Delta_x)^{m}\Omega^{(m)}_x$, resulting e.g. in the familiar $J^{(0)}_x=\Omega^{(0)}_x$ for the diffusive case.
To arrive at a differential equation, we consider the evolution of expectation values and go to the limit of long wavelengths ($\Delta_x \rightarrow \partial_x$) assuming large enough variation lengths in space, such that $\frac{d}{dt}\braket{\hat{S}_x^z}=(-\partial_x)^{m+1}\braket{\Omega^{(m)}_x}$.

To obtain a closed equation for the now coarse-grained charge density $\braket{\hat{S}_x^z}=\braket{\hat{S}_x^z}(t)$, we require a hydrodynamic assumption which relates the multipole current to the derivatives of the charge density (see e.g. Ref.~\cite{doyon2019lecture}). We therefore expand
$\braket{\Omega^{(m)}_x} = - D \, (\partial_x)^{l(m)} \braket{\hat{S}^z_x}$,
and our task is to find the \textit{lowest} possible (i.e. most scaling relevant) $l(m) \in \mathbb{N}$ such that  $D \neq 0$ is consistent with the conservation of all moments $Q^{(n\leq m)}$ (We provide a scaling analysis in the supplemental material that shows that non-linear terms in the expansion of $\braket{\Omega^{(m)}_x}$ are irrelevant). 

For charge-conserving interacting quantum systems ($m=0$), known to generically exhibit diffusive transport at late times~\cite{Mukerjee06, Lux14, Bohrdt16,leviatan2017quantum, Rakovszky18,Khemani18Hyd, Parker19, Gopalakrishnan19, Schuckert20}, we should obtain Fick's law $\braket{\Omega^{(0)}_x}= \braket{J^{(0)}_x} =-D\, \partial_x \braket{\hat{S}_x^z}$, i.e. $l(0)=1$, resulting in the usual diffusion equation for $\braket{\hat{S}^z_x}$. However, general solutions of the diffusion equation break higher-moment conservation. This is seen most easy for the example of a melting domain wall, which exhibits a net current of charge, violating dipole conservation. 

How can we thus generalize Fick's law to higher conserved moments?
We notice that in a closed system with open boundary conditions and in the absence of sinks or sources, the current $\braket{\Omega^{(m)}_x}_{eq.}= \braket{\Omega^{(m)}_x}(t\rightarrow \infty) =0$ is expected to vanish in equilibrium.
Combining this condition with $\braket{\Omega^{(m)}_x} = -D\, (\partial_x)^{l(m)} \braket{\hat{S}^z_x}$ leads to the equilibrium charge distribution
\begin{equation} \label{eq:a1.4}
\braket{\hat{S}^z_x}_{eq.} = a_0 + a_1\, x + ... + a_{l(m)-1}\, x^{l(m)-1} = \sum_{s=0}^{l(m)-1}a_s\, x^s.
\end{equation}
\eq{eq:a1.4} is a polynomial of degree $l(m)-1$ and contains a number $l(m)$ of independent constants $a_s$ which characterize the equilibrium state. On the other hand, since we assumed conservation of all moments $Q^{(n\leq m)}$, we know that equilibrium has to be characterized by $m+1$ independent parameters. This immediately determines $l(m) = m+1$, and the natural generalization of Fick's law is thus given by
$\braket{\Omega^{(m)}_x} = -D\, (\partial_x)^{m+1} \braket{\hat{S}^z_x}$.
This coincides with the intuition of the finite difference construction of $\Omega^{(m)}_x$: the dynamics balances out inhomogeneities of the $m^{th}$ derivative of the charge density.

Inserting this relation back into the evolution equation for the charge density, we finally arrive at the generalized hydrodynamic equation
\begin{equation} \label{eq:3.1}
\frac{d}{dt}\braket{\hat{S}^z_x} = -D(-1)^{m+1}(\partial_x)^{2(m+1)} \braket{\hat{S}^z_x},
\end{equation}
valid for systems conserving all multipole moments up to and including $m$. We emphasize that our derivation not only predicts the hydrodynamic equation \eq{eq:3.1}, but also the expected equilibrium distribution \eq{eq:a1.4} in closed systems, where the corresponding constants $a_s = a_s(Q^{(n\leq m)})$ are uniquely fixed by the charge moments $Q^{(n\leq m)}$ of the initial state. In systems of size $L$ they go as $|a_s| \sim \mathcal{O} (L^{-s})$, but are manifest in e.g. observables involving macroscopic distances like $\braket{\hat{S}^z_L}-\braket{\hat{S}^z_0}$. The prediction \eq{eq:a1.4} can be verified numerically in small systems by monitoring the charge distribution resulting from a fixed inital state at very late times. \figc{fig:3}{a} shows a chosen initial charge distribution in a system of size $L=20$, as well as the late time distributions obtained from evolving the system using both dipole- and quadrupole-conserving automata. The resulting distributions are in very good agreement with the predicted polynomials of \eq{eq:a1.4}, validating our approach.

We further notice that while usually, the hydrodynamic description of a system conserving $m+1$ quantities is given by a set of $m+1$ coupled equations for the associated densities and currents~\cite{chaikin_lubensky_1995,landau1987fluid,forster2018hydrodynamic}, the present systems are described by a \textit{single} equation \eq{eq:3.1} for the charge density. This is due to the hierarchical structure of the conservation laws \eq{eq:1} that specify all $Q^{(m)}$ in terms of the fundamental charges of the theory.

\textit{\textbf{Analytical Solution.}}--
It is worthwhile to consider the solutions of \eq{eq:3.1} in more detail. The normalized fundamental solution is of the form
\begin{equation} \label{eq:3.2}
 G^{(m)}(x,t)=\frac{1}{ (Dt)^{1\mathbin{/}2(m+1)}}F^{(m)}\Big[\frac{x^{2(m+1)}}{t}\Big],
\end{equation}
where $F^{(m)}$ is a universal scaling function which can be written in terms of generalized hypergeometric functions~\cite{Abramowitz65}. The time evolution of a charge density profile starting from $G^{(m)}(x,0)=\delta(x)$ is described by  \eq{eq:3.2}, and is thus expected to coincide with the correlator $C^{(m)}(x,t)$ by standard linear response theory~\cite{chaikin_lubensky_1995}. 
Figure \ref{fig:2} (c,d) shows that this is indeed the case, displaying full agreement with our numerical results for both $m=1,2$ upon fitting the only free parameter $D$. In particular, as demonstrated in \figc{fig:2}{c,d}, $C^{(m)}(x,t)$ accurately follows the scaling collapse predicted by \eq{eq:3.2}.

For $m=0$, \eq{eq:3.1} reduces to the usual diffusion equation and $C^{(0)}(x,t)$ is a Gaussian probability distribution describing the movement of an initially localized excitation through the system~\cite{chaikin_lubensky_1995,RevModPhys.54.195}. 
\begin{figure}[t]
\centering
\includegraphics[trim={0cm 0cm 0cm 0cm},clip,width=0.99\linewidth]{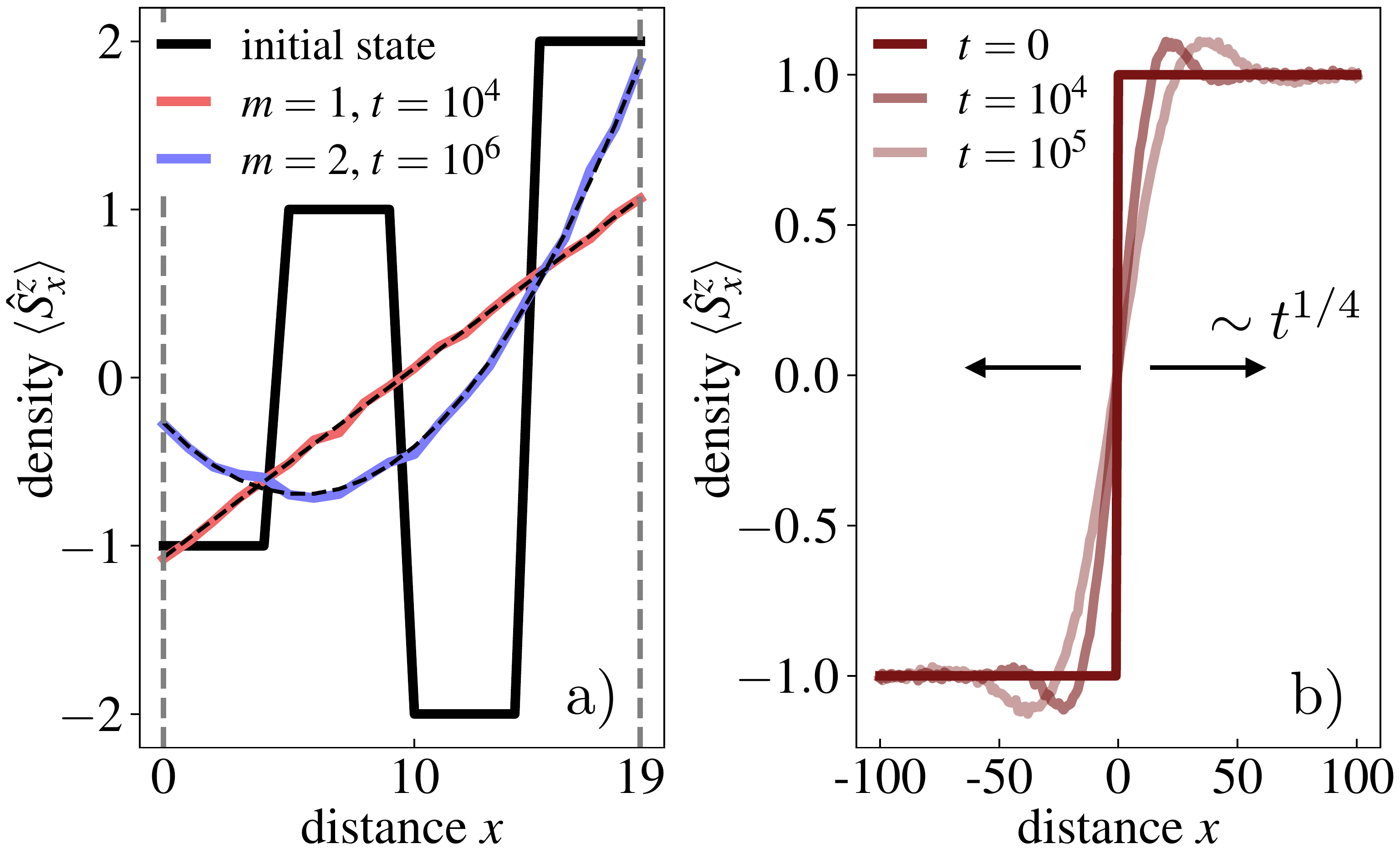}
\caption{\textbf{Implications of higher moment conservation.} \textbf{a)} In a finite size system with open boundary conditions (gray dashed lines), the charge density relaxes to an equilibrium distribution that is a polynomial of order $m$ (here: $S=3$). The black dashed lines are the analytical predictions from \eq{eq:a1.4}. \textbf{b)} The melting of a domain wall in a dipole conserving system for sufficiently large spin (here: $S=2$) appears as the cummulative distribution function of $C^{(1)}(x,t)$, with characteristic charge density oscillations.}
\label{fig:3}
\end{figure}
For $m \geq 1$, as shown in \fig{fig:2} and more generally clear from a vanishing second moment $\braket{x^2}_{G^{(m)}}=\int dx\, x^2 G^{(m)}(x,t)=0$, $C^{(m)}(x,t)$ \textit{cannot} be interpreted as a probability distribution. Instead, the associated oscillations in the profile of $C^{(m)}(x,t)$ form a characteristic signature of higher-moment conservation that can potentially also be observed in quench experiments of domain wall initial states, see \figc{fig:3}{b}.

Finally, we notice that the central peak of $C^{(m)}(x,t)$ in \figc{fig:2}{c,d} is well approximated by a Gaussian $g(x,t)=\exp\left(-x^2/\sigma^2(t)\right)/\sqrt{\pi\sigma^2(t)}$ with $\sigma(t) = (Dt)^{1/2(m+1)}$. The additional dressing density modulations can be understood heuristically if we interpret the Gaussian distribution $g(x,t)$ as describing the movement of an excitation through the system as part of $m$-poles. Conservation of $Q^{(m>0)}$ implies that a surrounding cloud of opposite charge has to be dragged along, see \figc{fig:1}{b}. The effective length scale for this process is given by $\sigma(t)$. This intuition can be formalized by making an Ansatz $C^{(m)}(x,t)=c_0\,g(x,t) - c_2\bigl[ g(x+\sigma(t)) +  g(x-\sigma(t)) \bigr] \approx g(x,t)[c_0-4c_2x^2/\sigma(t)^2]$. Thus, a positive charge moving to $x\pm \sigma(t)$ implies an increased likelihood of simultaneously finding a negative charge at $x$. Generalizing this physically motivated Ansatz, we can expand $C^{(m)}(x,t) = g(x,t)\,\sum_{n} c^{(m)}_{2n}\, \left(-\frac{x^2}{\sigma^2(t)}\right)^n$.
\figc{fig:2}{c,d} shows excellent agreement already at low orders of the expansion, where each term provides an additional oscillation in the spatial profile of $C^{(m)}(x,t)$.

\textit{\textbf{Conclusions \& Outlook.}}--
We have studied the long-time dynamics of higher-moment conserving models, obtaining a generalized hydrodynamic equation relevant for fractonic systems that leads to subdiffusive decay of charge correlations. We emphasize that for dipole-conservation, our results provide a postdiction of the subdiffusive scaling experimentally observed in Ref.~\cite{Guardado20}, where an initially prepared $k$-wave density mode of interacting fermions was found to decay as $\sim \exp(-k^4t)$ in the presence of a strong, linearly tilted potential. The linear potential couples directly to the center of mass $\sum_x x\,\hat{n}_x$ (see also~\cite{Mandt2011_interacting}) and can thus be thought of as inducing an effective dipole conservation on long length scales. The observed decay agrees with \eq{eq:3.1} when written in Fourier space. In analogy, the present analysis suggests that effective quadrupole-conservation may be obtained by application of a \textit{harmonic} potential.

In addition to the subdiffusive decay of the return probability, we have identified oscillations in the spatial density profile both for delta and domain wall initial conditions as characteristic properties of higher-moment conservation. Such oscillations should be detectable in quantum quench experiments. Furthermore, higher-moment conservation leads to a modified scaling of the full-width-half-maximum of the Lorentzian line shape as $\sim k^{2(m+1)}$ in Fourier space, which could be detectable in scattering experiments~\cite{minkiewicz1969_critical}. 
Finally, we expect our results to reflect the long-time dynamics of closed quantum systems as indicated by investigations on small system sizes, see the supplementary material. Understanding the full impact of higher moment conservation laws on the dynamics of quantum systems, in particular in the presence of energy conservation, remains an objective for future study.

\textit{\textbf{Acknowledgments.}}-- We thank Monika Aidelsburger, Alexander Schuckert, and Adam Smith for insightful discussions. We especially thank Tibor Rakovszky for discussions and for carefully reading the manuscript.
We acknowledge support from the Technical University of Munich - Institute for Advanced Study, funded by the German Excellence Initiative and the European Union FP7 under grant agreement 291763, the Deutsche Forschungsgemeinschaft (DFG, German Research Foundation) under Germany’s Excellence Strategy--EXC--2111--390814868,  “la Caixa”Foundation (ID 100010434) fellowship grant for post-graduate studies,  Research Unit FOR 1807 through grants no. PO 1370/2-1, TRR80 and DFG grant No. KN1254/1-1, No. KN1254/1-2, and from the European Research Council (ERC) under the European Union’s Horizon 2020 research and innovation programme (grant agreements No. 771537 and No. 851161).

\textit{\textbf{Note added:}} While finalizing this manuscript, we became aware of a related work by  A. Gromov, A. Lucas and R. M. Nandkishore~\cite{gromov2020_fractonhydro} and a related work by A. Morningstar, V. Khemani and D. Huse which appeared in the previous arXiv posting~\cite{morningstar2020_kinetic}.


\bibliography{Subdiff_1D}

\newpage
\leavevmode \newpage
\onecolumngrid
\begin{center}
\textbf{Supplemental Material:}\\
\textbf{Anomalous Diffusion in Dipole- and Higher-Moment Conserving Systems}\\ \vspace{10pt}
Johannes Feldmeier$^{1,2}$, Pablo Sala$^{1,2}$, Giuseppe De Tomasi$^{3}$, Frank Pollmann$^{1,2}$, and Michael Knap$^{1,2}$ \\ \vspace{6pt}

$^1$\textit{\small{Department of Physics and Institute for Advanced Study, Technical University of Munich, 85748 Garching, Germany}} \\
$^2$\textit{\small{Munich Center for Quantum Science and Technology (MCQST), Schellingstr. 4, D-80799 M{\"u}nchen, Germany}} \\
$^3$\textit{\small{T.C.M. Group, Cavendish Laboratory, JJ Thomson Avenue, Cambridge CB3 0HE, United Kingdom}}
\vspace{10pt}
\end{center}
\maketitle
\twocolumngrid

\section{Quantum model with Dipole conservation}

In order to further support our main results, in this section, we show the  dynamics of a quantum model featuring dipole conservation. For numerical feasability, we choose $S=1/2$ and study the Hamiltonian given by

\begin{equation}
\label{eq:H_quantum}
 \hat H = \hat H_4^{(1)} + \hat H_5^{(1)},
\end{equation}
on open boundary conditions, where
\begin{equation}
 \hat H_4^{(1)} = - \sum_{x=1}^{L-3} [S_x^{+} S_{x+1}^{-} S_{x+2}^{-} S_{x+3}^+ + h.c.],
\end{equation}
and
\begin{equation}
 \hat H_5^{(1)} = - \sum_{x=1}^{L-4} [S_x^{+} S_{x+1}^{-} S_{x+3}^{-} S_{x+4}^+ + h.c.],
\end{equation}
$L$ indicates the length of the chain. It is readily verified that $[\hat H, \sum_x \hat S_x^z] = 0$ and $[\hat H, \sum_x x \hat S_x^z] = 0$, thus $\hat H$ conserves the total magnetization and dipole moment. We focus on the largest Hamiltonian sector with $\langle \sum_x \hat S^z_x \rangle = \langle  \sum_x x \hat S_x^z\rangle  = 0$.
Moreover, the Hamiltonian $\hat H$ in Eq.~\ref{eq:H_quantum} has been already analyzed in Refs.~\cite{Taylor_19, Sala19},  showing that $\hat H$ is ergodic and its Hilbert space is only weakly fragmented.

We focus on the dynamics of $\hat H$ by employing the spin-spin correlator, also defined in the main text

\begin{equation}
\label{eq:correlator}
 C^{(1)}(t) = \langle \delta \hat  S_{L/2}^z(t) \delta \hat S_{L/2}^z(0) \rangle,
 \end{equation}
 where $\langle \cdot \rangle = \frac{1}{\mathcal{N}}\text{Tr}[\cdot ]$ is the normalized infinite-temperature trace, with $\mathcal{N}$ the dimension of the Hilbert space sector $\langle \sum_x \hat S^z_x \rangle = \langle  \sum_x x \hat S_x^z\rangle  = 0$, and $\delta \hat  S_{L/2}^z(t) = \hat  S_{L/2}^z(t) - \langle \hat  S_{L/2}^z(t) \rangle$.

\begin{figure}[t!]
\includegraphics[width=1.\columnwidth]{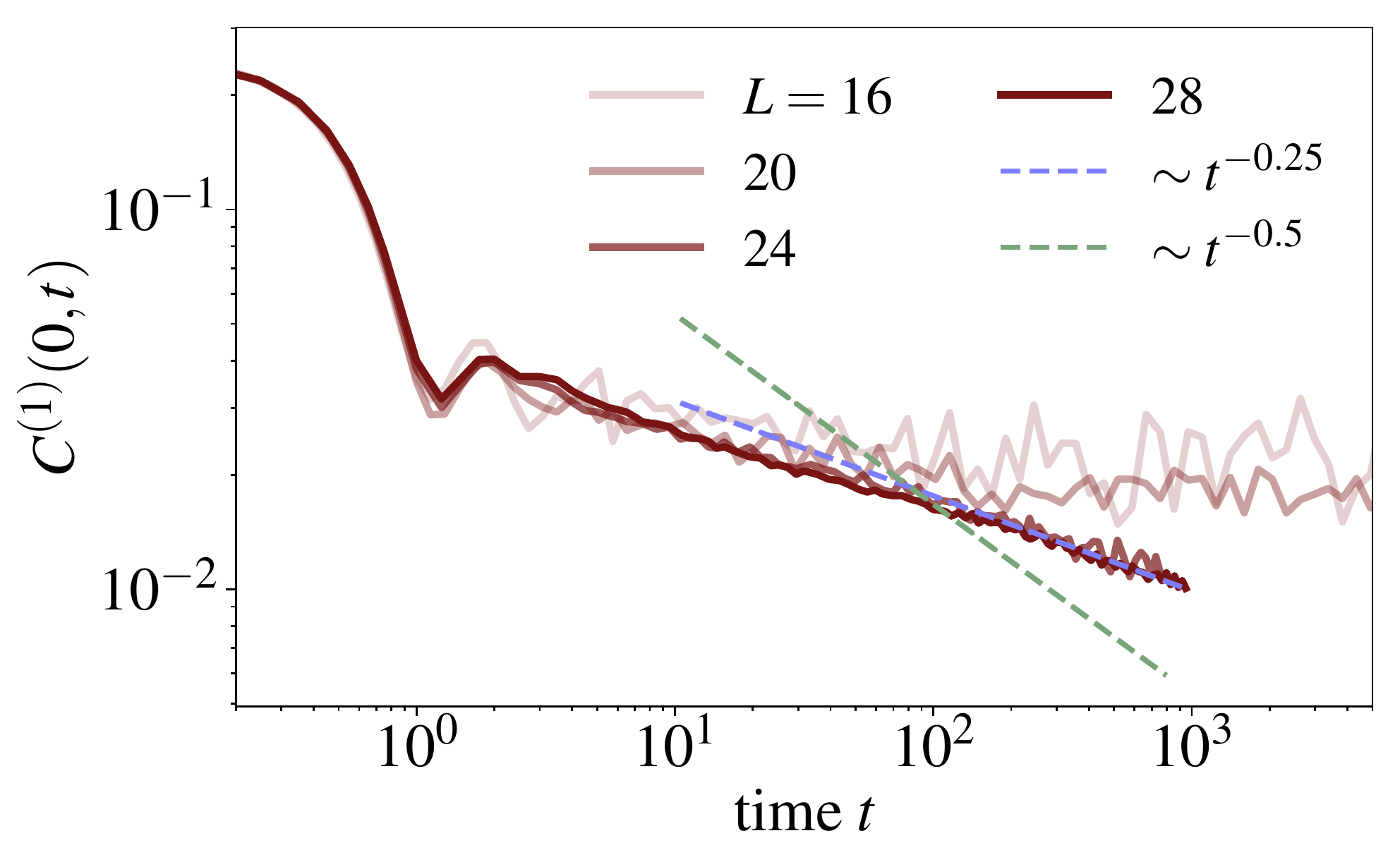}
\caption{Return probability $C^{(1)}(t)$, defined in Eq.~\ref{eq:correlator}, for several $L \in \{16,20,24,28\}$. $C^{(1)}(t)$ decays to zero algebraically $C^{(1)}(t) \sim t^{-\alpha}$, with $\alpha$ consistent with a subdiffusive relaxation $\alpha \approx 1/4$. The dashed-lines $\sim t^{-0.5}$ and $\sim t^{-0.25}$ are guides for the eye.}
\label{fig:Fig4_quantum}
\end{figure}

Figure~\ref{fig:Fig4_quantum} shows $C^{(1)}(t)$ for several system sizes $L \in \{16,20,24,28\}$. 
For small system sizes $L\in \{16,20\}$, $C^{(1)}(t)$ has been computed using exact diagonalization, and for $L\in \{24,28\}$ using Chebyshev polynomials techniques for the time evolution where the trace has been evaluated stochastically~\cite{weisse_kernel,bera2017_density}. As expected,  after a short time propagation ($t\sim O(1)$), $C^{(1)}(t)$ relaxes algebraically to zero, $C^{(1)}(t) \sim t^{-\alpha}$. This relaxation to the equilibrium value is subdiffusive, $\alpha < 1/2$. In agreement with the result in the main text, the observed decay appears to be consistent with the subdiffusive exponent $\alpha = 1/4$ for ergodic dipole conserving Hamiltonians (blue dash-line in Fig.~\ref{fig:Fig4_quantum}).

\section{Automaton Time Evolution}
We provide additional details on the stochastic automaton scheme employed in the main text. As touched upon therein, the automaton evolution can be expressed in terms of a classically simulable circuit where the updates are represented as unitary operators $\hat{U}^{(m)}_r(x)$ of range $r$, acting on local strings of $z$-basis configurations $\ket{\bs{s}(x)}=\ket{s_x,...,s_{x+r-1}}$ and conserving all $Q^{(n\leq m)}$. These operators connect two states $\ket{\bs{s}(x)}$ and $\ket{\bs{s}^\prime(x)}=\hat{U}^{(m)}_r(x)\ket{\bs{s}(x)}$, such that $\hat{U}^{(m)}_r(x)\ket{\bs{s}^\prime(x)}=\ket{\bs{s}(x)}$~\cite{Iaconis19}, i.e.,  $\hat{U}^{(m)}_r(x)$ acts as a usual  $\sigma^x$ Pauli matrix in the subspace spanned by $\ket{\bs{s}(x)}$ and $\ket{\bs{s}^\prime(x)}$. For states that are locally frozen due to the boundaris of the chosen spin representation, $\hat{U}^{(m)}_r(x)$ acts as the identity. For a selected $h^{(m)}_r(x)$ of Eq.~(2) of the main text, there are potentially \textit{two} possible updates for a given $\ket{\bs{s}(x)}$, namely $\ket{\bs{s}^\prime(x)}=h^{(m)}_r(x)\ket{\bs{s}(x)}$ and $\ket{\bs{s}^{\prime\prime}(x)}= \left(h^{(m)}_r(x)\right)^\dagger\ket{\bs{s}(x)}$. Therefore, we require \textit{two} independent unitary gates $\hat{U}^{(m)}_{r,A}(x)$ and $\hat{U}^{(m)}_{r,B}(x)$ describing all possible updates associated to a local term of $\hat{H}^{(m)}_r$. These update operators are the building blocks of the automaton evolution. The definition of a single time step in terms of these gates for the case $m=1$, taking into account only $\hat{H}^{(1)}_3$, is illustrated in \fig{fig:2}.

We can further provide a more explicit construction of the two local unitaries $\hat{U}^{(m)}_{r,A/B}(x)$ in the following way: let us define the elementary unitary operators $\hat{U}_{\bs{s},\bs{s}^\prime}$, which act as $\hat{U}_{\bs{s},\bs{s}^\prime}\ket{\bs{s}(x)} = \ket{\bs{s}^\prime (x)}$ and $\hat{U}_{\bs{s},\bs{s}^\prime}\ket{\bs{s}^\prime (x)} = \ket{\bs{s}(x)}$, and as the identity on all states outside the two-state subspace spanned by $\{\ket{\bs{s}(x)},\ket{\bs{s}^\prime (x)}\}$. We can define such an elementary unitary for every pair $(\bs{s}(x),\bs{s}^\prime (x))$ of local spin configurations that are connected by the Hamiltonian $\hat{H}^{(m)}_r$. Because in general, as described above, a given $\ket{\bs{s}}$ can be connected to two other configurations by $\hat{H}^{(m)}_r$, we can seperate the set of all possible pairs $(\bs{s}(x),\bs{s}^\prime (x))$ into two subsets $A$ and $B$, such that a given $\bs{s}(x)$ appears only at most \textit{once} as part of a pair $(\bs{s}(x),\bs{s}^\prime (x))$ in $A$, and analogously in $B$. We can then define the abovementioned $\hat{U}^{(m)}_{r,A/B}(x)$ in terms of the elementary unitaries acting only on pairs within one of the two sets $A$ and $B$ via
\begin{equation} \label{eq:A2.1}
\hat{U}^{(m)}_{r,A} = \prod_{(\bs{s},\bs{s}^\prime) \in A} \hat{U}_{\bs{s},\bs{s}^\prime}
\end{equation}
and analogously for $\hat{U}^{(m)}_{r,B}$. Due to the construction of the sets $A$ and $B$, the unitaries in the product of \eq{eq:A2.1} all commute.
 
\begin{figure}[t!]
\includegraphics[width=0.8\columnwidth]{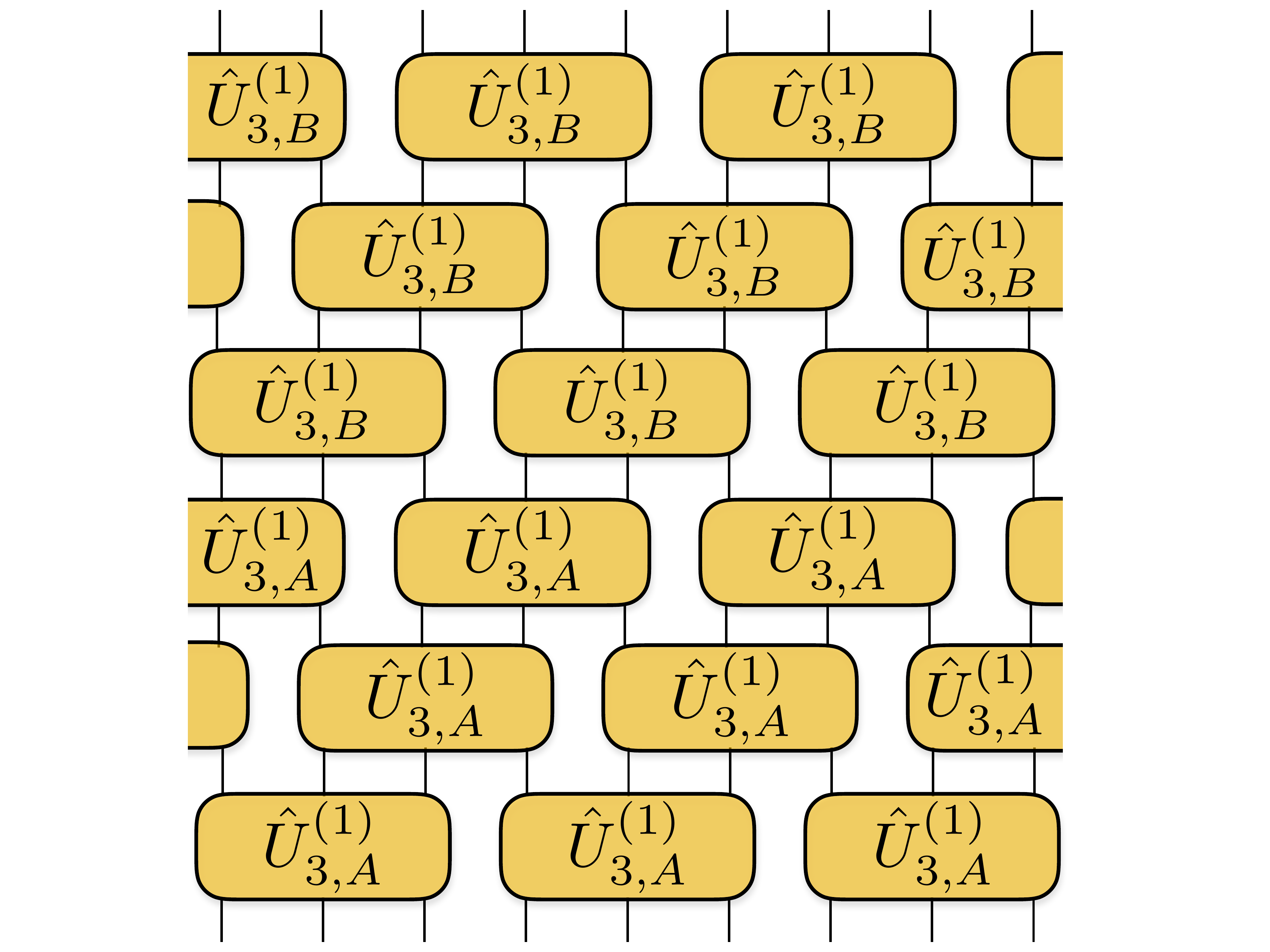}
\caption{Single time step of the automaton circuit evolution using the unitary operators defined in the text.}
\label{fig:2}
\end{figure}

The relation of the circuit dynamics to the Hamiltonian $\hat{H}^{(m)}_r$ can then be expanded even further by endowing each application of an update between $\ket{\bs{s}(x)}$ and $\ket{\bs{s}^\prime(x)}$ with an acceptance probability $p \sim \left|\bra{\bs{s}^\prime(x)}(h^{(m)}_r(x)+h.c.)\ket{\bs{s}(x)}\right|^2$ proportional to the associated Fermi-golden-rule rate. According to this probability, the update gate is either applied or substituted by an identity operator. This turns our circuit into a \textit{stochastic} automaton~\cite{Ritort03} (cf. main text).\\

\section{Scaling analysis}
We provide a general scaling analysis that demonstrates the validity of the expansion for the multipole current employed in the main text. Let us start with the hydrodynamic equation for the coarse-grained charge density of a system conserving all $Q^{(n\leq m)}$ as derived in the main text,
\begin{equation} \label{eq:a4.1}
\begin{split}
&\frac{d}{dt} \braket{\hat{S}^z_x} - (-\partial_x)^{m+1} \braket{\Omega^{(m)}_x} =\\ 
&\quad = \frac{d}{dt} \braket{\hat{S}^z_x} + D\, (-1)^{m+1}(\partial_x)^{2(m+1)} \braket{\hat{S}^z_x} = 0.
\end{split}
\end{equation}
Here, we have used the expansion $\braket{\Omega^{(m)}_x} \approx -D\, (\partial_x)^{m+1}\braket{\hat{S}^z_x}$ discussed above. Our goal is to determine the relevance of taking into account additional terms $\braket{\Omega^{(m)}_x} = \braket{\Omega^{(m)}_x}\bigl( \partial_x^j \braket{S^z_x}^i  \bigr)$ with $i,j \in \mathbb{N}$ in a more general expansion of the coarse-grained multipole current.

We do so following the scheme laid out in Ref.~\cite{Lux14,lux2016_thesis} for diffusive systems, by first extending \eq{eq:a4.1} to include microscopic noise fluctuations $\xi(x,t)$ of the multipole current, i.e. $\braket{\Omega^{(m)}_x} \rightarrow \braket{\Omega^{(m)}_x} + \xi(x,t)$, such that
\begin{equation} \label{eq:a4.2}
\frac{d}{dt} \braket{\hat{S}^z_x} + D\, (-1)^{m+1}(\partial_x)^{2(m+1)} \braket{\hat{S}^z_x} = (-\partial_x)^{m+1}\xi(x,t)
\end{equation}
with uncorrelated noise
\begin{equation} \label{eq:a4.3}
\braket{\xi(x,t)\, \xi(x^\prime,t^\prime)} = \eta\, \delta(x-x^\prime)\, \delta(t-t^\prime).
\end{equation}
The amplitude of the noise is tied to the correlation functions of the charge density in equilibrium~\cite{Lux14,lux2016_thesis}, and is of microscopic origin. \eq{eq:a4.2} and \eq{eq:a4.3} are the starting points of our scaling approach: 
Consider a scale transformation $x \rightarrow x/\lambda$ with $\lambda > 1$. The scaling dimension of $x$ is thus given by $[x]= -1$. Demanding that \eq{eq:a4.2} be a fixed point under this rescaling, i.e. $[D]=0$, then fixes the scaling dimensions of $t,\xi, \braket{\hat{S}^z_x}$: Firstly, we obtain $[t] = -2(m+1)$ from the left hand side of \eq{eq:a4.2}. Second, \eq{eq:a4.3} implies $[\xi] = m+3/2$ (in one spatial dimension; note $[\eta]=0$ due to its connection to fluctuations in equilibrium~\cite{lux2016_thesis}). Finally, \eq{eq:a4.2} yields $[\braket{\hat{S}^z_x}]=1/2$. Given these scaling dimensions, we can assess whether additional terms included in the expansion of $\braket{\Omega^{(m)}_x}$ are relevant or not, and subsequently, whether they are consistent with multipole conservation laws.\\

We see that including terms $\braket{\Omega^{(m)}_x} \sim \alpha_{i,j}\, (\partial_x)^j\braket{S^z_x}^i$ in \eq{eq:a4.1} lead to a scaling dimension $[\alpha_{i,j}] = m+1-j + \frac{1}{2}\, (1-i)$ of the corresponding coefficient $\alpha_{i,j}$. In particular, all terms with $j>m+1,\, i\geq 1$ as well as $j = m+1,\, i > 1$ are irrelevant under the RG flow.\\

How about potentially relevant/marginal terms, which have $j\leq m+1$ and  $1\leq i \leq 2(m+1-j)+1$? For $j=m+1$ we are led back to the term used in \eq{eq:a4.1}. All other relevant/marginal terms necessarily have $j<m+1$, and thus \textit{fewer} derivatives than the number of $m+1$ independent constants necessary to characterize equilibrium. Following the argument used in the main text, integrating the associated condition $\braket{\Omega^{(m)}_x}_{eq.} = 0$ of a vanishing multipole current in equilibrium does hence not provide sufficiently many freely adjustable parameters for the equilibrium charge distribution. In other words, for terms with $j < m+1$ it is always possible to find initial states such that the conservation of some $Q^{(n\leq m)}$ is broken.

Overall, we indeed find that the expansion $\braket{\Omega^{(m)}_x} \approx -D\, (\partial_x)^{m+1}\braket{\hat{S}^z_x}$ used in the main text captures the unique, most scaling relevant term consistent with all multipole conservation laws.\\

\end{document}